\documentclass[onehalfspace,jeeabib]{jeea}

\usepackage[ruled,vlined,linesnumbered]{algorithm2e}
\usepackage{subfiles}
\usepackage{indentfirst,latexsym,bm}
\usepackage{booktabs}
\usepackage{bbm}
\usepackage{xcolor}
\usepackage{amsfonts}
\usepackage{amsmath,amsfonts} 
\usepackage{graphicx}
\usepackage{hyperref}
\usepackage[switch,pagewise]{lineno}
\usepackage{graphicx}

\Title{Testing Goodness-of-Fit for Conditional Distributions: A New Perspective based on Principal Component Analysis}
\Date{}
\Author{Rui}{Cui}{\\}{cuirui.econ@outlook.com}
\Author{Yuhao}{Li}{Xi'an Jiaotong-Liverpool University}{yuhao.li@xjtlu.edu.cn}
\Runninghead{Cui and Li}{ }
\Editor{}
\Thanks{We are deeply indebted to Miguel A. Delgado for his advice and support at all stages of this paper. The first author is particularly grateful to Winfried Stute for his guidance and care during the stay in Giessen. We also thank: Carlos Velasco, Juan Carlos Escanciano, Jes\'us Gonzalo, Nazarii Salish, Junyi Peng Zhou, Julius Vainora, Minghai Mao, Xiaojun Song. Li acknowledges support from the Research Development Fun (RDF-23-02-022) of the Xi'an Jiaotong-Liverpool University. All errors belong to us.}

\Abstract{
    This paper introduces a novel goodness-of-fit test technique for parametric conditional distributions. The proposed tests are based on a residual marked empirical process, for which we develop a conditional Principal Component Analysis. The obtained components provide a basis for various types of new tests in addition to the omnibus one. Component tests that based on each component serve as experts in detecting certain directions. Smooth tests that assemble a few components are also of great use in practice. To further improve testing performance, we introduce a component selection approach, aiming to identify the most contributory components. The finite sample performance of the proposed tests is illustrated through Monte Carlo experiments.}

\begin{document}
\maketitle
\section{Introduction}
In regression analysis, conditional distributions play a fundamental role in constructing economic models. They fully characterize the relationship between a response variable and its covariates while enabling insights of crucial statistical indicators—including expectations, variance, quantiles, and hazard rates. These distributions form the core of numerous econometric frameworks, from categorical or discrete choice models (e.g., (ordered) logit/probit) to count data regression models (Poisson, negative binomial, hurdle, and zero-inflation models) and duration analyses (Cox proportional hazards and accelerated failure time models). In these settings, the conditional distribution is typically specified up to a finite-dimensional parameter. The validity of estimation and inference procedures crucially depends on the correct specification of this parametric form, making model checking essential for their reliable use.

Let $Y$ be a real-valued response variable and $X$ a vector of covariates. This paper investigates specification tests for assessing whether the conditional distribution of $Y$ given $X$ belongs to a parametric family indexed by an unknown parameter vector $\theta$. Formally, we consider testing
\begin{equation*}
    \label{null_hypothesis}
    H_0: \mathbb{P}\left[Y \leq y \mid X\right] = F\left(y \mid X, \theta_0\right), \quad \text{a.s. for all } y \in \mathbb{R} \text{ and some } \theta_0 \in \Theta,
\end{equation*}
against its negation. The parameter space $\Theta$ is assumed to be a subset of $\mathbb{R}^p$, where $p$ is a fixed positive integer.

Although there is an extensive literature on consistent testing of conditional expectations—see, for example, \cite{bierens1982consistent,bierens1990consistent,bierens1997asymptotic,stute1997nonparametric,stinchcombe1998consistent,delgado2001significance,fan2000consistent,escanciano2006consistent,hardle1993comparing,horowitz1994testing,hong1995consistent,zheng1996consistent,lavergne2000nonparametric}—the literature on consistent specification testing for conditional distributions is comparatively limited. Only a few papers have addressed this problem, including \cite{andrews1997conditional,zheng2000consistent,zheng2012testing,bierens2012integrated,amengual2020testing}. Among these, \cite{andrews1997conditional} and \cite{bierens2012integrated} employ a similar methodology, which is based on transforming the conditional moment restriction into a continuum of unconditional moment conditions. This methodology is commonly known as the Integrated Conditional Moment (ICM) approach. In parallel, kernel-based tests have been developed by \cite{zheng2000consistent,zheng2012testing}, which compare the parametric model to nonparametric or semiparametric estimates using kernel smoothing techniques.

Compared to kernel-based tests, a key advantage of the ICM approach is its ability to address the curse of dimensionality in covariates. Kernel-based tests are particularly vulnerable to the curse of dimensionality, as they require nonparametric kernel density estimation, which becomes increasingly unreliable in high dimensions. In contrast, the ICM framework naturally accommodates dimension reduction techniques. For instance, \cite{escanciano2006consistent} proposes using projection weights based on indicator functions in the transformed unconditional moments, while \cite{bierens1982consistent} advocates for characteristic function weighting. 

However, the performance of ICM tests are still unsatisfactory as they often exhibit limited power in practice. While theoretically consistent—being omnibus tests capable of detecting all possible deviations—they generally lack detection power. A more critical limitation is that although these tests can identify general model misspecification, they fail to reveal which specific aspects of the data lead to rejection. This naturally raises an important question: Can the omnibus test be decomposed into components, each capturing a distinct aspect of the data? If possible, evaluating both the overall test and the significance of individual components would provide more informative diagnostics and deeper insights into the nature of detected deviations.

The idea of decomposing omnibus tests into orthogonal components originates from the work of \cite{durbin1972components}, who introduced Principal Component Analysis (PCA) for the classical empirical process in the context of unconditional distribution goodness-of-fit testing. Through functional PCA, they achieved an orthogonal decomposition of the omnibus test statistic into principal components (PCs), with each PC capturing a distinct direction of departure from the null hypothesis. This approach was later extended by \cite{durbin1975components} to composite hypothesis testing with estimated parameters. Building on these foundations, \cite{stute1997nonparametric} further demonstrated the significant application of PCA in goodness-of-fit tests by discussing the issue of checking the mean regression model. He derived the PCs of the residual marked process, which serve as a basis for constructing more efficient tests, such as Neyman-type smooth tests and optimal directional tests.

While PCA is a powerful tool for enhancing test power, only a handful of studies have successfully derived the PCs of omnibus test statistics in explicit form. Most existing work merely establishes the existence of an orthogonal decomposition without providing its concrete representation. Following the foundational work of Durbin and Stute, we develop an extension of PCA application for general conditional distribution testing. However, previous PCA approaches—designed for univariate stochastic processes—prove inadequate in conditional distribution case. The introduction of covariates in distribution functions fundamentally alters the empirical process's structure, transforming it from univariate to multivariate. The resulting complex dependence structure in the multivariate process presents the primary challenge for implementing PCA in this setting.

To address this challenge, we propose a novel methodology called conditional Principal Component Decomposition (PCD). Our approach demonstrates that the residual-marked empirical process can be decomposed into a series of asymptotically independent component processes. Each component process consists of individual PCs and serves as the foundation for our new class of specification tests. Notably, the dimension reduction strategies developed for ICM tests can be seamlessly incorporated into our PCD framework, enabling effective handling of high-dimensional covariates.

PCA also provides a straightforward explanation for the underperformance of omnibus tests. In omnibus statistics, the contribution of each PC is weighted by its corresponding eigenvalue, which typically decays rapidly. As a result, higher-order components—those most sensitive to high-frequency departures from the null—are heavily downweighted and thus contribute little to the overall test statistic. This makes omnibus tests inherently less effective at detecting such high-frequency alternatives. By contrast, once the PCs are explicitly identified, we gain the flexibility to assign weights as desired: individual components can be used as targeted tests for specific types of deviations, while reweighted combinations (smooth tests) can be constructed to enhance power against a broader class of alternatives. In this sense, PCs serve as fundamental building blocks for more powerful and informative tests.

Despite this flexibility, a practical challenge remains: there is little theoretical guidance on how to optimally select and combine components for a given dataset. Traditionally, researchers have relied on the heuristic that the first few components capture most of the relevant departures from the null, but this approach may be unreliable in complex settings—such as those involving high-dimensional covariates—where the signal may be spread across many components. Consequently, smooth tests that simply aggregate the leading components can exhibit substantial variation in power. This underscores the need for data-driven methods to identify and utilize the most informative components, thereby making full use of the diagnostic potential offered by PCA.

In this paper, we proposes a component selection procedure based on a ``learning before testing'' strategy using sample splitting. In the learning sample, we estimate the p-values of individual component tests to gauge their significance, selecting those with the smallest p-values as the most informative. These selected components are then aggregated—typically with equal weights—into a smooth test statistic in the testing sample. Our simulation studies demonstrate that even this straightforward approach can substantially enhance testing power. Further research on optimal weighting schemes for the selected components is left for future work.

The remainder of the paper is organized as follows. In Section 2, after a brief reviewing of existing omnibus ICM tests for conditional distributions, we introduce our conditional PCD methodology and present the construction of component-based test statistics, and establishes their asymptotic properties. Section 3 describes the proposed component selection procedure. Section 4 investigates the finite-sample performance of the new tests through Monte Carlo simulations. We also apply the proposed methodology to a real data example in this section. Section 5 concludes.

\section{Tests based on Principal Components}
\subsection{Omnibus ICM Tests}

Consider a sample $\{Y_i, X_i\}_{i=1}^n$ consisting of i.i.d. realizations of the random vector $(Y, X)$, where $Y$ is the real-valued response variable and $X$ is the $d$-dimensional covariate vector. Our objective is to assess the conditional distribution of $Y$ given $X$. To test $H_0$, the ICM approach introduces a suitable weighting function $h(v, X)$, allowing the conditional moment condition
$$ \mathbb{E}\left[\boldsymbol{1}_{\{Y \leq y\}} - F(y|X, \theta_0) \mid X\right] = 0, \quad \text{a.s.} $$
to be equivalently reformulated as a continuum of unconditional moment conditions:
$$ \mathbb{E}\left[\left(\boldsymbol{1}_{\{Y \leq y\}} - F(y|X, \theta_0)\right) h(v, X)\right] = 0, \quad \forall v \in \Pi, $$
for some index set $\Pi$. Several families of weighting functions satisfy this equivalence, with the most common choices being the characteristic function (see \cite{bierens1982consistent}) and the indicator function (see \cite{andrews1997conditional, stute1997nonparametric}). While the literature often uses $x$ as the argument to denote the weighting function's independent variable, we use $v$ to distinguish it from the actual values of the covariates, which will be important in later sections. We assume $v$ is $d'$-dimensional.

To test the unconditional moment conditions, which is considerably easier, we introduce the empirical process
$$R_{n}(y,v):=n^{-1/2}\sum_{i=1}^{n}h(v,X_i) M_i(y), \eqno{(2.1)}$$
where
$${M}_i(y)=\boldsymbol{1}_{\{Y_i\le y\}} - F \left(y | X_i, {\theta}\right). \eqno{(2.2)}$$
Here, the ``residual'' process $M_i$ is written separately, as it will play a central role in our subsequent decomposition. By analyzing the behavior of the ${R}_n$ process, we can make statistical inferences: if the process deviates sufficiently from zero, we reject the null hypothesis. 

In fact, all existing ICM tests for conditional distributions are fundamentally based on the $R_{n}$ process, with the primary distinction being the choice of weighting function $h$. For instance, if $h$ is selected as the indicator function $\boldsymbol{1}_{\{X \leq v\}}$ with $v \in \mathbb{R}^d$, we obtain the process introduced by Andrews (1997):
$${\alpha}_{n}(y,v) := n^{-1/2} \sum_{i=1}^{n} \boldsymbol{1}_{\{X_i \leq v\}} M_i(y). \eqno{(2.3)}$$
Its empirical version 
$$\hat{\alpha}_n(y,v) := n^{-1/2} \sum_{i=1}^{n} \boldsymbol{1}_{\{X_i \leq v\}} \hat{M}_i(y),$$
where
$$\hat{M}_i(y) = \boldsymbol{1}_{\{Y_i \leq y\}} - F\big(y \mid X_i, \hat{\theta}\big),$$
and $\hat{\theta}$ is a consistent estimator of $\theta,$ is used to built Andrews' conditional Kolmogorov-Smirnov (CK) test, namely,
$$CK = \sup_{y, v} \big| \hat{\alpha}_n(y, v) \big|. \eqno{(2.4)}$$

However, Andrews' test suffers from the curse of dimensionality due to the use of the indicator function. To address this, \cite{escanciano2006consistent} proposed replacing the weighting function $\boldsymbol{1}_{\{X\le v\}}$ with projection weights, which mitigates the high-dimensionality problem. Although originally developed for regression models, this approach is also applicable to conditional distributions. Specifically, the empirical process becomes
$$\hat{\zeta}_{n}(\beta,z,y)=n^{-1/2}\sum_{i=1}^{n}\boldsymbol{1}_{\{\beta^\top X_i\le z\}} \hat{M}_i(y),$$
where the weighting function is $\boldsymbol{1}_{\{\beta^\top X\le z\}}$ with $v=(\beta;z)\in \mathbb{R}^{d+1}$. A Cramér-von Mises (CvM) type statistic can be defined as
$$CvM_{es}=\int \big(\hat{\zeta}_{n}(\beta,z,y)\big)^2  F_{n, Y}(dy) F_{n,\beta}(dz)d\beta, \eqno{(2.5)}$$
where $F_{n, Y}(y)$ and $F_{n,\beta}(z)$ are the empirical distributions of $Y$ and $\beta^\top X$, respectively, and $d\beta$ denotes the uniform measure on the unit sphere.

Another widely used weighting function, introduced by \cite{bierens1982consistent}, is the characteristic function $\exp(i v^{\top} X)$ with $v \in \mathbb{R}^d$. In the context of conditional distributions, the corresponding CvM statistic is
$$CvM_{ch}=\int\big(\hat{\eta}(y,v)\big)^2  F_{n,Y}(dy)  \Upsilon(v)dv,   \eqno{(2.6)}$$
where
$$ \hat{\eta}(y,v) =  n^{-1/2}\sum_{i=1}^{n}\exp(i v^{\top} X_i) \hat{M}_i(y), $$
and $\Upsilon(v)$ is the standard multivariate normal density.

All these tests share the ICM framework and are consistent against all possible alternatives, but the choice of weighting function can significantly affect their power performance. Escanciano (2006) argues that his projection-based test combines the advantages of both Andrews' and Bierens' approaches. It preserves the convenience of having a natural integration measure for CvM statistics offered by the indicator weighting function, and also has the ability to handle high-dimensional problems associated with exponential weighting. 

Another concern of ICM tests lies in their low power against high-frequency alternatives. As \cite{fan2000consistent} demonstrated, these tests can be viewed as kernel tests with a fixed bandwidth. In such cases, the fixed bandwidth may oversmooth the conditional distribution, masking important features of the alternative.

In this paper, we address the power limitations of omnibus tests for high-frequency alternatives from a different perspective. Rather than focusing solely on the choice of weighting function, we develop a functional PCA for the $R_{n}$ process, decomposing it into an infinite series of asymptotically independent component processes. Each component process captures deviations in a specific frequency direction, providing a foundation for constructing more powerful tests. Importantly, this decomposition is independent of the choice of $h$, thereby preserving the same flexibility in selecting weighting functions for the component processes as in the original  $R_{n}$. As a result, our approach enables effective handling of both high-dimensional covariates and high-frequency alternatives. The details of the decomposition are presented in the next section.

Before proceeding, we introduce the key assumptions regarding the conditional distribution model and the estimation procedure.

\noindent\textbf{(A1).} The function $F(y| X_i,\theta)$ is differentiable with respect to $\theta$ in a neighborhood of $\theta_0$ for all $i \geq 1$.

\begin{flalign*}
    \noindent\textbf{(A2).} \quad \sup_{(y,v)\in \mathbb{R}^{d'+1}} \sup_{\theta : \|\theta-\theta_0\|\le r_n}\Bigg\| \frac{1}{n}\sum_{i=1}^{n}\frac{\partial}{\partial \theta} F(y| X_i,\theta) h(v,X_i) -\Delta_0(y,v)  \Bigg\| \to 0, \quad a.s.
\end{flalign*}
for all sequences of positive constants $\{r_n : n\ge 1\}$ with $r_n\to 0$, where $\Delta_0(y,v)=\int (\partial / \partial \theta)F(y| s,\theta_0)h(v,s)\,dF_X(s)$ and $F_X$ denotes the marginal distribution of $X$.

\noindent\textbf{(A3).} $\sup_{(y,v)\in \mathbb{R}^{d'+1}} \| \Delta_0(y,v) \| <\infty$
and $\Delta_0(\cdot)$ is uniformly continuous on $\mathbb{R}^{d'+1}$.

\noindent\textbf{(A4).} The parametric estimator admits the expansion
$$
n^{1/2}\big(\hat{\theta}-\theta_0\big) = n^{-1/2}\sum_{i=1}^{n}l(X_i,Y_i,\theta_0) + o_p(1),
$$
for some function $l$ satisfying $\mathbb{E}[l(X,Y,\theta_0)] = 0$, and $L(\theta_0) = \mathbb{E}\left[l(X,Y,\theta_0)l^{\top}(X,Y,\theta_0)\right]$ exists and is positive definite.

Assumptions (A1)-(A3) are standard in the literature; see, for example, \cite{andrews1997conditional}. Assumption (A4) requires that the maximum likelihood estimator satisfies the usual regularity conditions, which is also commonly imposed; see, for instance, \cite{bierens1982consistent,bierens1997asymptotic,bierens2012integrated,delgado2001significance}.


 \subsection {Conditional Principal Component Decomposition}
Since the process $R_n$ is multivariate and involves dependent components $y$ and $v$, it does not admit a direct Karhunen-Loève representation. Nevertheless, thanks to the martingale structure of each ${M}_i$, a frequency-domain decomposition is still possible. Our conditional PCD approach leverages this fact that each ${M}_i$ is a centered process with a Brownian Bridge type covariance structure, hence admits an explicit PCD. The spectral decomposition of $R_n$ is then obtained by aggregating the PCs of all ${M}_i$ processes. Before presenting the details, we introduce some necessary notation.

Let
$$\mu_j = \frac{1}{(\pi j)^2}, \qquad \varphi_j(y) = \sqrt{2}\sin(j\pi y), \quad j=1,2,\ldots \eqno{(2.7)}$$
be the eigenvalues and eigenfunctions of the standard Brownian Bridge $B(t)$, whose covariance kernel is $K(s,t) = s \wedge t - st$, with $s \wedge t = \min(s,t)$. For each $x$, introduce the transformation
$$T(y, x) := F(y \mid X = x, \theta_0),  \eqno{(2.8)}$$
where $\theta_0$ is the true parameter value. The function $T$ is non-decreasing in $y$, with $T(-\infty, x) = 0$ and $T(\infty, x) = 1$. Applying this transformation to $\varphi_j$, we define compound functions
 $$f_j(y,x):=\varphi_j(T(y,x)), \quad j=1,2,\ldots$$
They are actually the eigenfunctions of transformed Brownian Bridge $B(T(y,x))$ with fixed $x$. Also, their cosine counterpart are important, thus we define
 $$
g_j(y,x):=\sqrt{2}\cos(j\pi T(y,x)).$$

For each fixed $x$, the collection $\{f_j(\cdot, x)\}_{j=1}^\infty$ forms an orthonormal basis for a subspace of $L^2(\mathbb{R}, T(\cdot, x))$, the Hilbert space of square-integrable functions on $\mathbb{R}$ with an inner product
$$\langle \rho, g \rangle_x = \int_{\mathbb{R}} \rho(y) g(y) T(dy, x).$$
Indeed,
\begin{eqnarray*}
    \langle f_j, f_h \rangle_x &=& \int_{\mathbb{R}} \varphi_j(T(y, x)) \varphi_h(T(y, x)) T(dy, x) \\
    &=& \int_0^1 \varphi_j(u) \varphi_h(u) du = \begin{cases} 1 & j = h \\ 0 & j \neq h \end{cases}.
\end{eqnarray*}
Moreover, $\{f_j(\cdot, x)\}_{j=1}^\infty$ are the eigenfunctions of $M(y)=\boldsymbol{1}_{\{Y \leq y\}} - F\big(y \mid X, {\theta_0}\big)$ conditional on $X = x$, with corresponding eigenvalues $\{\mu_j\}_{j=1}^\infty$. This result follows from the conditional covariance kernel of $M(y)$:
\begin{eqnarray*}
    \mathbb{E}(M(y_1) M(y_2) \mid X = x) &=& F(y_1 \wedge y_2 \mid X = x, \theta_0) - F(y_1 \mid X = x, \theta_0) F(y_2 \mid X = x, \theta_0) \\
    &=& K(T(y_1, x), T(y_2, x)).
\end{eqnarray*}
It is identical to the covariance kernel of $B(T(y,x))$, therefore the eigenfunctions of $M(y)$ are given by the transformation of $\varphi_j$.

By Mercer's theorem, the covariance kernel of $M(y)$ admits the following spectral decomposition:
$$K(T(y_1, x), T(y_2, x)) = \sum_{j=1}^{\infty} \mu_j f_j(y_1, x) f_j(y_2, x). \eqno{(2.9)}$$
The PCD of each $M_i$ is then straightforward. By substituting all these PCDs into $R_n$, we obtain a decomposition of $R_n$ itself. This result is formalized in the following proposition, with the proof provided in the appendix.\\

\begin{scshape}
    \noindent\textbf{Proposition 1} 
\end{scshape}
\begin{itshape}
    Under the null hypothesis, the process $(2.1)$ can be represented as a weighted sum of component processes:
    $$R_n(y,v) = \sum_{j=1}^{\infty} \mu_j^{1/2} c_{n,j}(y,v), \eqno{(2.10)}$$
    where the weights are $\mu_j^{1/2} = (\pi j)^{-1}$, and the component processes are given by     
    $$c_{n,j}(y,v) = n^{-1/2} \sum_{i=1}^{n} h(v, X_i) f_j(y, X_i) g_j(Y_i, X_i), \quad j = 1, 2, \ldots$$ 
\end{itshape}

We now focus on the component processes $c_{n,j}$, as they will form the basis for our proposed tests. Each $c_{n,j}$ is a sum of i.i.d. terms involving the weighting function $h(v, X_i)$, the eigenfunction $f_j(y, X_i)$, and a random variable $g_j(Y_i, X_i)$. The variable $g_j(Y_i, X_i)$ is the $j$-th normalized principal component of $M_i$ conditional on $X_i$, and thus possesses the standard orthogonality properties of principal components, but conditional on $X_i$. Specifically, for each $j$ and $j \neq h$,
\begin{align*}
    & \mathbb{E}[g_j(Y_i, X_i) \mid X_i] = 0, \qquad \mathbb{E}[g_j^2(Y_i, X_i) \mid X_i] = 1,\\
    & \mathbb{E}[g_j(Y_i, X_i) g_h(Y_i, X_i) \mid X_i] = 0.
\end{align*}

The i.i.d. structure of the component processes $c_{n,j}$ is also useful for deriving their asymptotic properties. Each $c_{n,j}$ is a sum of i.i.d. centered random functions with variance
$$
H_j(y,v) := \mathbb{E}\left[ h^2(v,X) f_j^2(y,X) \right] = \int h^2(v,s) f_j^2(y,s) F_X(ds),
$$
where $F_X(\cdot)$ denotes the distribution function of $X$. The following theorem describes their asymptotic behavior.

\bigskip

\begin{scshape}
    \noindent\textbf{Theorem 1}
\end{scshape}
\begin{itshape}
    Under the null hypothesis and assumptions $(A1)$--$(A3)$, for each $j$, the process $c_{n,j}(y,v)$ converges weakly to a centered Gaussian process,
    $$
    c_{n,j} \stackrel{d}{\to} c_{\infty,j}.
    $$
    The limiting process $c_{\infty,j}(y,v)$ has covariance function
    $$
    K(y_1, y_2, v_1, v_2) = \int h(v_1, s) h(v_2, s) f_j(y_1, s) f_j(y_2, s) F_X(ds).
    $$
    Moreover, $c_{\infty,j}$ and $c_{\infty,h}$ are independent for $j \neq h$.
\end{itshape}


\subsection {Tests based on component processes}
The component processes derived above form the foundation for a new class of specification tests for conditional distributions. Our proposed tests are based on their empirical counterparts, defined as
$$
\hat{c}_{n,j}(y,v) := n^{-1/2} \sum_{i=1}^{n} h(v, X_i) \hat{f}_j(y, X_i) \hat{g}_j(Y_i, X_i). \eqno{(2.11)}
$$
Here,
$\hat{f}_j(y, x) = \sqrt{2} \sin\big(j\pi \hat{T}(y, x)\big)$ and $\hat{g}_j(y, x) = \sqrt{2} \cos\big(j\pi \hat{T}(y, x)\big)$,
where $\hat{T}(y, x)$ is an estimator of $T(y, x)$. A natural and consistent choice is
$$
\hat{T}(y, x) = F\big(y \mid X = x, \hat{\theta}\big).
$$

Given a specific weighting function $h$ in (2.11), we can construct CvM type statistics for each component process. Specifically, we introduce our component test statistic as
$$
CvM_{n,j} = \int \left(\hat{c}_{n,j}(y, v)\right)^2 \Psi_{n}(dy, dv), ~~j = 1, 2, \ldots,  \eqno{(2.12)}
$$
where $\Psi_{n}(dy, dv)$ is the empirical measure determined by the choice of $h$ in $\hat{c}_{n,j}$. For example, if Escanciano's projection weight $\boldsymbol{1}_{\{\beta^\top X_i \leq z\}}$ is used, $\Psi_{n}$ corresponds to $F_{n, Y}(dy) F_{n, \beta}(dz) d\beta$ as in (2.5); if Bierens' characteristic function $\exp(i v^{\top} X_i)$ is used, it becomes $F_{n, Y}(dy) \Upsilon(v) dv$ as in (2.6).

Beyond individual component tests, one can construct test statistics by combining several component processes. Specifically, consider the first $m$ component processes, reweighted by a sequence $w = (w_1, w_2, \ldots, w_m)$. The corresponding CvM type statistic\footnote{Analogously, KS type statistics can be defined as $$KS_{n,j} = \sup_{y,v} \big| \hat{c}_{n,j}(y,v) \big|$$ and $$KS_{n,m}^w = \sup_{y,v} \left| \sum_{j=1}^{m} w_j \hat{c}_{n,j}(y,v) \right|.$$} is given by
$$
CvM_{n,m}^w = \int \left( \sum_{j=1}^{m} w_j \hat{c}_{n,j}(y,v) \right)^2 \Psi_{n}(dy, dv). \eqno{(2.13)}
$$
A smooth test, in the spirit of Neyman's smooth test, can be constructed by setting $w = (1, \ldots, 1)$, i.e., by equally weighting the first $m$ components.  Smooth tests offer a balance between omnibus tests and individual component tests, and are often effective against a broad class of alternatives. However, the choice of the weight vector $w$ is critical  in practical applications, an issue we will address in later section.

We finish this part by providing the asymptotic theories of the estimated component processes and their reweighted combinations. Formal proofs are in the Appendix.\\

\begin{scshape}
    \noindent\textbf{Theorem 2}
\end{scshape}
\begin{itshape}
    Under the null hypothesis and assumptions $(A1)$--$(A4)$, for each $j=1, 2, \ldots$, the process $\hat{c}_{n,j}(y,v)$ converges weakly to a centered Gaussian process,
    $$
    \hat{c}_{n,j} \stackrel{d}{\to}  \tilde{c}_{\infty,j}.
    $$
    The limiting Gaussian process $\tilde{c}_{\infty,j}(y,v)$ has covariance structure  
    \begin{align*} 
        K(y_1, y_2, v_1, v_2)  =
        \mathrm{cov}  \Big[ & h(v_1, X) {f}_j(y_1, X) {g}_j(Y, X) -  A_j(y_1, v_1) l(X, Y, \theta_0), \\
        & h(v_2, X) {f}_j(y_2, X) {g}_j(Y, X) -  A_j(y_2, v_2) l(X, Y, \theta_0) \Big],
    \end{align*}
    where
    $$
        A_j(y,v)={\mu_j}^{-1/2}\mathbb{E}\left[h(v,X){f}_j(y,X)\int_{\mathbb{R}}F_{\theta}(y|X, \theta_0)f_j(y,X)T(dy,X)  \right].
    $$
\end{itshape}

For the reweighted process $\sum_{j=1}^{m} w_j \hat{c}_{n,j}(y,v)$, the asymptotic distribution coincides with that of $\sum_{j=1}^{m} w_j \tilde{c}_{n,j}(y,v)$. Its limiting behavior follows directly.\\

\begin{scshape}
    \noindent\textbf{Theorem 3}
\end{scshape}
\begin{itshape}
    Under the null hypothesis and assumptions $(A1)$--$(A4)$, for any given weight vector $w = \{w_j\}_{j=1}^{m}$, the process $\sum_{j=1}^{m} w_j \hat{c}_{n,j}(y,v)$ converges weakly to a centered Gaussian process,
    $$
    \sum_{j=1}^{m} w_j \hat{c}_{n,j} \stackrel{d}{\to}  \tilde{c}_{\infty,m}^{w}.
    $$
    The limiting Gaussian process $\tilde{c}_{\infty,m}^{w}(y, v)$ has covariance structure  
    \begin{align*}
        K(y_1, y_2, v_1, v_2)  =
        \mathrm{cov} \Bigg[ \sum_{j=1}^{m} w_j \Big( h(v_1, X) {f}_j(y_1, X) {g}_j(Y, X) -  A_j(y_1, v_1) l(X, Y, \theta_0) \Big), \\
        \sum_{j=1}^{m} w_j \Big( h(v_2, X) {f}_j(y_2, X) {g}_j(Y, X) -  A_j(y_2, v_2) l(X, Y, \theta_0) \Big) \Bigg],
    \end{align*}
    where $A_j(y,v)$ is defined as in Theorem 2. 
\end{itshape}

The convergence of our CvM test statistics then follows immediately by the continuous mapping theorem. In practice, following \cite{andrews1997conditional}, we estimate the critical values of these tests using a parametric bootstrap procedure.

A crucial remark that need to be made is that our proposed tests—including the component tests $CvM_{n,j}$ and the smooth tests $CvM_{n,m}^w$—are generally not consistent, as they are only parts of the omnibus tests in certain spectral directions. However, in practice they often outperform the omnibus tests.  As equation (2.10) shows, the weight $(\pi j)^{-1}$ assigned to the $j$-th component process decays rapidly with increasing $j.$ Consequently, higher-order components contribute progressively less to $R_n$, making the omnibus test less sensitive to deviations captured by these components. Importantly, these higher-order components detect higher-frequency departures, which explains the omnibus test's limited power against high-frequency alternatives. In contrast, individual component tests are specifically designed to detect such particular types of deviations. To demonstrate this specialized detection capability, we present an example in the following section that illustrates how different components serve as targeted detectors for specific types of departures from the null.

 \subsection {Role of components: a simple simulation example}
In the context of unconditional normal distributions, \cite{durbin1972components} observed that different PCs are sensitive to departures in specific moments: the first PC is most responsive to mean shifts, the second to variance, the third to skewness, and the fourth to kurtosis. This insight led to the recommendation that each PC be examined individually to better understand the nature of any detected deviation.

To investigate whether this correspondence between PCs and moments extends to conditional distributions, we consider a conditional normal model with a linear conditional mean and constant variance:
$$
H_0: Y \mid X \sim \mathcal{N}(\theta_0 + \theta_1 X, \theta_2),
$$
where $\theta = (\theta_0, \theta_1, \theta_2)$. For simplicity, we let $X$ be real-valued and construct several data generating processes (DGPs) that introduce deviations in mean, variance, skewness, and kurtosis, respectively.

The first two DGPs are specified as follows:
\begin{itemize}
    \item \textbf{Mean shift:} $Y \sim \mathcal{N}(1 + X + 5\cos(2X), 1)$;
    \item \textbf{Heteroscedasticity:} $Y \sim \mathcal{N}(1 + X, 6(X-0.5)^2 + 0.5)$.
\end{itemize}

For skewness and kurtosis, we use a conditional distribution function of the form
$$
F(y \mid x, \gamma_3, \gamma_4) = \Phi(y - 1 - x) + \gamma_3 \sin(3\pi \Phi(y - 1 - x)) + \gamma_4 \sin(4\pi \Phi(y - 1 - x)),
$$
where $\Phi(\cdot)$ denotes the standard normal cumulative distribution function. This construction, adapted from (8.4) in \cite{durbin1975components}, allows $\gamma_3$ and $\gamma_4$ to control the degree of skewness and kurtosis, respectively:
\begin{itemize}
    \item \textbf{Skewness shift:} $\gamma_3 = 0.1$, $\gamma_4 = 0$;
    \item \textbf{Kurtosis shift:} $\gamma_3 = 0$, $\gamma_4 = 0.1$.
\end{itemize}

We evaluate the performance of the first four component tests, using Andrews' indicator weighting function (appropriate for the univariate covariate case). Critical values are obtained via standard parametric bootstrap, as detailed in Section 5. Table \ref{tab:a1} reports the power of each component test, with the highest power for each scenario highlighted in bold.

\begin{table}[!htp]
	\centering
	\caption{Component Test Results, Deviations in the first four moments}
	\setlength{\tabcolsep}{18pt}
	\label{tab:a1}
	\resizebox{\textwidth}{!}{
		\scriptsize
		\renewcommand{\arraystretch}{1.5} 
		
		\begin{tabular}{@{}lllllllllllll@{}}
			\toprule
			& \multicolumn{3}{c}{Mean deviation} && \multicolumn{3}{c}{Variance} \\
			\cmidrule(lr){2-4} \cmidrule(lr){6-8}
			& $n=100$ & $200$ & $300$  && $n=100$ & $200$ & $300$ \\
			\midrule
			$CvM_{n,1}$         & \bf{0.444} & \bf{0.780} & \bf{0.904} && 0.194 & 0.222 & 0.178 \\
			\addlinespace 
			$CvM_{n,2}$         & 0.082 & 0.118 & 0.114 && \bf{0.264} & \bf{0.426} & \bf{0.476} \\
			\addlinespace
			$CvM_{n,3}$         & 0.042 & 0.064 & 0.066 && 0.006 & 0.006 & 0.002 \\
			\addlinespace
			$CvM_{n,4}$         & 0.074 & 0.066 & 0.084 && 0.048 & 0.050 & 0.056 \\
			\addlinespace
			\bottomrule

& \multicolumn{3}{c}{Skewness} && \multicolumn{3}{c}{Kurtosis} \\
\cmidrule(lr){2-4} \cmidrule(lr){6-8}
& $n=100$ & $200$ & $300$  && $n=100$ & $200$ & $300$ \\
\midrule
$CvM_{n,1}$         & 0.618 & 0.906 & 0.968 && 0.024 & 0.030 & 0.028 \\
\addlinespace 
$CvM_{n,2}$         & 0.104 & 0.162 & 0.198 && 0.078 & 0.110 & 0.098 \\
\addlinespace
$CvM_{n,3}$         & \bf{0.838} & \bf{0.992} & \bf{1.000} && 0.054 & 0.070 & 0.068 \\
\addlinespace
$CvM_{n,4}$         & 0.118 & 0.178 & 0.192 && \bf{0.226} & \bf{0.292} & \bf{0.360} \\
\addlinespace
\bottomrule
		\end{tabular}
	}
	\begin{tablenotes}
		$CvM_{n,j}$, $j=1,2,\ldots,4$ are the $ j$-th component test with $ n $ being the sample size.  
	\end{tablenotes}
\end{table}

We observe the same relationship pattern between PCs and moments in the conditional normal case as that recorded in \cite{durbin1972components} for the unconditional setting. The first four component tests are each most sensitive to deviations in the first four moments, respectively. These component tests serve as experts for detecting moment departures from conditional normality. 

Moreover, in many practical situations, deviations tend to be concentrated in the first few components, making smooth tests that assemble these leading components particularly effective. However, this experience and the above pattern may fail in high-dimensional settings—even when using projection weights (Escanciano) or characteristic function weights (Bierens). In such cases, the first few components may not be the most informative, and the clear correspondence between PCs and moments, as seen in the normal case, can break down. For example, Table \ref{tab:c3} in Appendix B shows that for two high-dimensional normal DGPs with different mean shift scales, the third and eighth components are most powerful for detecting mean deviations.

 Consequently, when prior knowledge about the roles of PCs is unreliable, identifying which specific components capture the most significant deviations in a given dataset becomes crucial, as this can substantially improve testing efficiency. To address this challenge, we introduce a ``learning then testing'' procedure in the next section, which provides a data-driven way to identify the most informative components, thereby guiding the construction of more efficient tests.

\section{Component Selection}
In this section, we present a straightforward, data-driven component selection strategy. The idea is to split the full sample into two parts: a learning sample and a testing sample. In the learning sample, we estimate the p-values for each component test by applying a parametric bootstrap to approximate the empirical null distribution. The components are then ranked according to their p-values, and the $m$ components with the smallest p-values are selected for further analysis. These selected components are subsequently used to construct the test statistic in the testing sample. The following pseudo-code outlines this procedure.

\begin{algorithm}
\SetAlgoLined
\caption{Learning the P-Values}
\label{alg:1}
    \textbf{Input:} $Z= (X, Y) $, Bootstrap sample size $ B$, component number $ M$ \;
    Split the sample into learning and testing parts: $ \{X_{learn}, Y_{learn}\}, \{X_{test},Y_{test}\} $\;
    
    Using the learning sample, obtain a consistent estimator $ \hat{\theta}_{learn}$, as well as $ M$ component test statistics $ \{\hat{c}_{j}(\hat{\theta}_{learn})\}_{j\in [M]}$   \;

    Perform parametric bootstrap in the learning sample for each component process and obtain the associated empirical distribution $ \{\hat{c}_{j,b}\}_{b \in [B]} $ \; 

    Calculate p-values: $ p_{j} = mean(\hat{c}_{j}(\hat{\theta}_{learn}) \le \{\hat{c}_{j,b}\}_{b \in [B]} )$ \;

    Sort these p-values in ascend order, select the first $ m$ components, record the corresponding indexes and form a vector $ \mathcal{W}$  \;

\end{algorithm}

\begin{algorithm}
\SetAlgoLined
\caption{Testing Procedures}
\label{alg:2}
    \textbf{Input:} $Z_{test}= (X_{test}, Y_{test}) $, Bootstrap sample size $ B$, a vector of component index $ \mathcal{W}$ learned from Algorithm \ref{alg:1}, significance level $ \alpha$  \;
    
    Using the testing sample, obtain a consistent estimator $ \hat{\theta}_{test}$, as well as component test statistics $ \{\hat{c}_{j}(\hat{\theta}_{test})\}_{j\in \mathcal{W}}$   \;
    
    Construct an equal weight smooth test statistic $ \bar{c} = mean( \{\hat{c}_{j}(\hat{\theta}_{test})\}_{j\in \mathcal{W}})$  \;

    Perform parametric bootstrap in the testing sample for each component process in $ \mathcal{W}$ and obtain the associated bootstrap statistic $ \{\hat{c}_{j,b}\}_{j \in \mathcal{W},b \in [B]}$ \;
    
    Obtain the empirical distribution for the smooth statistic: $ \{\bar{c}_b\}_{b \in [B]}$, where $ \bar{c}_b = mean(\{\hat{c}_{j,b}\}_{j \in \mathcal{W}})$ \;

    Calculate the p-value $p= mean(\bar{c}\le \{\bar{c}_b\}_{b \in [B]})$, reject the null if $ p<\alpha$.   
\end{algorithm}

During the learning procedure, it is important to assess the stability of the selected components, particularly the variability of the estimated p-values. Several strategies can be employed to enhance the robustness of the learning process. A straightforward approach is to increase the size of the learning sub-sample, which may reduce the variance of the p-values. However, this comes at the cost of a smaller testing sub-sample, potentially diminishing the power of the final test. Determining the optimal split between learning and testing samples is a nontrivial problem and is left for future research.

Another strategy is to use bagging. This involves repeatedly resampling the learning sub-sample and applying Algorithm \ref{alg:1} to obtain multiple sets of p-values for each component. Averaging these p-values across resamples can reduce their variability and improve the reliability of component selection. The main drawback of this approach is the increased computational burden.

In the testing procedure, a key unresolved issue is the choice of weights assigned to each component in the smooth test statistic. In this paper, we adopt the simple approach of assigning equal weights to all selected components. Ideally, components with smaller p-values—indicating stronger evidence against the null—should receive larger weights, so that their influence on the test statistic is amplified. However, the choice of weights also affects the variance of the test statistic, and inappropriate weighting may actually reduce power. At present, there is limited theoretical guidance on optimal weight selection, and further research is needed to address this question.

\section{Simulation and Empirical Studies}
\noindent 
\subsection{Simulation Studies}
In this section, we evaluate the performance of our proposed test statistics using Monte Carlo simulations. The data-generating processes (DGPs) considered are as follows:

\begin{itemize}
    \item DGP0: $ Y \sim \mathcal{N}(1+ \beta^\top X,1) $;
    \item DGP1: $ Y \sim \mathcal{N}(1+\beta^\top X,(2\beta^\top(X-0.5))^2) $;
    \item DGP2: $Y \sim \mathrm{Exp}(1+ \beta^\top X)$;
    \item DGP3-5: $Y \sim \mathcal{N}(1+\beta^{\top}X+5 \cos(k\beta^{\top}X),3.5)$, with $k=3,4,5$;
    \item DGP6-8: $Y \sim \mathcal{N}(1+ 5 \cos(k\beta^{\top}X),(2\beta^\top(X-0.5))^2)$, with $k=5,6,7$;
\end{itemize}

For all DGPs, the covariates $X$ are generated independently from the uniform distribution $U(0,1)$, with dimension $d=15$. All coefficients are set to one, i.e., $\beta = (1,\ldots,1)^\top$.

DGP1 examines deviations in both mean and variance, while DGP2 introduces a distributional deviation. DGPs 3--5 focus on mean deviations with increasing frequency, and DGPs 6--8 are designed to illustrate the benefits of the learning-based procedure in detecting more complex alternatives.

We compare the proposed smooth test, both with and without the learning-then-testing procedure. The characteristic function is used as the weighting function $h$. The default smooth test, denoted $Smooth\_m$, is constructed by equally weighting the first $m$ components from the full sample. Alternatively, $Smooth\_Learn\_m$ is computed in the testing sub-sample, using the $m$ components identified in the learning phase as most indicative of deviation. In both cases, components are equally weighted.

For benchmarking, we also report results for three omnibus test statistics: (i) Andrew's test ($CK$), (ii) the CvM test with Escanciano's projection weighting ($CvM_{es}$), and (iii) the CvM test with characteristic function weighting ($CvM_{ch}$), as defined in equations (2.4)--(2.6).

Table \ref{tab:0} reports the estimated empirical size at the 95\% confidence level for sample sizes ranging from 100 to 300. The results indicate that the proposed component-based tests, including those using the learning procedure, exhibit only minor size distortions, which diminish as the sample size increases.

\begin{table}[!htp]
    \centering
    \caption{Estimated size for DGP0 at 95\%}
    \setlength{\tabcolsep}{18pt}
    \label{tab:0}
    \resizebox{\textwidth}{!}{
        \scriptsize
        \renewcommand{\arraystretch}{1.5} 
        
        \begin{tabular}{@{}lllllllllllll@{}}
            \toprule
            & \multicolumn{3}{c}{DGP0}  \\
            \cmidrule(lr){2-4} 
            & $n=100$ & $200$ & $300$   \\
            \midrule
            $CK$         & 0.046 & 0.058 & 0.046  \\
            \addlinespace
            $CvM_{ch}$         & 0.088 & 0.068 & 0.052  \\
            \addlinespace
            $CvM_{es}$ & 0.074 & 0.044 & 0.056  \\
            \addlinespace
            Smooth\_5           & 0.088 & 0.052 & 0.056\\
            \addlinespace
            Smooth\_10          & 0.08 & 0.062 &0.062 \\
            \addlinespace
            Smooth\_{Learn}\_5  & N/A & 0.068 & 0.056 \\
            \addlinespace
            \bottomrule
        \end{tabular}
    }
    \begin{tablenotes}
        The Smooth\_m is built by equally weighing the first $m$ components. For the learning process, the first 50 observations are used. The Smooth\_Learn\_m considers the learned largest $m$ components with a sample size of $ n-50 $. Unless otherwise noted, this applies to all the tables below.
    \end{tablenotes}
\end{table}

\begin{table}[!htp]
    \centering
    \caption{Estimated power for DGP1-8 at 95\%}
    \setlength{\tabcolsep}{18pt}
    \label{tab:1}
    \resizebox{\textwidth}{!}{
        \scriptsize
        \renewcommand{\arraystretch}{1.5} 
        
        \begin{tabular}{@{}lllllllllllll@{}}
            \toprule
            & \multicolumn{3}{c}{DGP1} && \multicolumn{3}{c}{DGP2} \\
            \cmidrule(lr){2-4} \cmidrule(lr){6-8}
            & $n=100$ & $200$ & $300$  && $n=100$ & $200$ & $300$ \\
            \midrule
            $CK$         & 0.598 & 0.446 & 0.258  && 0.018 & 0.038 & 0.048  \\
            \addlinespace
            $CvM_{ch}$         & 0.978 & 0.994 & 0.988  && 1.000 & 1.000 & 1.000\\
            \addlinespace
            $CvM_{es}$ & 0.754 & 0.818 & 0.846 && 1.000 & 1.000 & 1.000\\
            \addlinespace 
            Smooth\_5           & 0.966 & 0.998 & 1.000 && 0.774 & 0.870 & 0.916\\
            \addlinespace
            Smooth\_10          & 0.984 & 1.000 &1.000  && 0.574 & 0.860 &0.962\\
            \addlinespace
            Smooth\_{Learn}\_5  & N/A & 0.992 & 1.000   && N/A & 0.962 & 0.976\\
            \addlinespace
            \bottomrule

            & \multicolumn{3}{c}{DGP3} && \multicolumn{3}{c}{DGP4} \\
            \cmidrule(lr){2-4} \cmidrule(lr){6-8}
            & $n=100$ & $200$ & $300$  && $n=100$ & $200$ & $300$ \\
            \midrule
            $CK$         & 0.008 & 0.034 & 0.058 && 0.016 & 0.048 & 0.050 \\
            \addlinespace
            $CvM_{ch}$         & 0.052 & 0.080 & 0.054 && 0.066 & 0.036 & 0.034 \\
            \addlinespace
            $CvM_{es}$ & 0.046 & 0.054 & 0.048 && 0.042 & 0.028 & 0.026\\
            \addlinespace 
            Smooth\_5           & 0.092 & 0.216 & 0.216 && 0.094 & 0.170 & 0.238\\
            \addlinespace
            Smooth\_10          & 0.098 & 0.174 &0.188  && 0.088 & 0.142 &0.194\\
            \addlinespace
            Smooth\_{Learn}\_5  & N/A & 0.192 & 0.312   && N/A & 0.196 & 0.300  \\
            \addlinespace
            \bottomrule

            & \multicolumn{3}{c}{DGP5} && \multicolumn{3}{c}{DGP6} \\
            \cmidrule(lr){2-4} \cmidrule(lr){6-8}
            & $n=100$ & $200$ & $300$  && $n=100$ & $200$ & $300$ \\
            \midrule
            $CK$         & 0.022 & 0.036 & 0.056  && 0.572 & 0.394 & 0.260\\
            \addlinespace
            $CvM_{ch}$         & 0.064 & 0.046 & 0.046  && 0.888 & 0.914 & 0.948\\
            \addlinespace
            $CvM_{es}$ & 0.040 & 0.030 & 0.030  && 0.570 & 0.636 & 0.656\\
            \addlinespace 
            Smooth\_5           & 0.108 & 0.192 & 0.244  && 0.412 & 0.454 & 0.536\\
            \addlinespace
            Smooth\_10          & 0.092 & 0.144 &0.164   && 0.284 & 0.376 &0.518\\
            \addlinespace
            Smooth\_{Learn}\_5  & N/A & 0.180 & 0.280    && N/A & 0.606 & 0.630\\
            \addlinespace
            \bottomrule

            & \multicolumn{3}{c}{DGP7} && \multicolumn{3}{c}{DGP8} \\
            \cmidrule(lr){2-4} \cmidrule(lr){6-8}
            & $n=100$ & $200$ & $300$  && $n=100$ & $200$ & $300$ \\
            \midrule
            $CK$         & 0.510 & 0.382 & 0.264  && 0.536 & 0.412 & 0.240\\
            \addlinespace
            $CvM_{ch}$         & 0.856 & 0.910 & 0.920  && 0.906 & 0.918 & 0.932 \\
            \addlinespace
            $CvM_{es}$ & 0.532 & 0.628 & 0.688  && 0.512 & 0.630 & 0.644\\
            \addlinespace 
            Smooth\_5           & 0.362 & 0.448 & 0.488  && 0.430 & 0.422 & 0.494\\
            \addlinespace
            Smooth\_10          & 0.250 & 0.370 &0.466   && 0.278 & 0.348 &0.488\\
            \addlinespace
            Smooth\_{Learn}\_5  & N/A & 0.640 & 0.620    && N/A & 0.640 & 0.598\\
            \addlinespace
            \bottomrule
        \end{tabular}
    }

\end{table}

Table \ref{tab:1} summarizes the power results, from which several key observations emerge. 

First, component-based smooth tests generally outperform omnibus tests, but their effectiveness depends critically on the choice of component weights. For instance, in DGP3-5, all three omnibus tests ($CK$, $CvM_{ch}$, and $CvM_{es}$) exhibit negligible power, whereas the smooth tests—whether constructed from the first 5 or 10 components or using the learning procedure—demonstrate substantially higher power. In contrast, for DGP6-8, the omnibus tests ($CvM_{ch}$ and $CvM_{es}$) outperform the smooth tests, which show relatively poor power. This pattern can be explained by examining the power of individual component tests (see Appendix Tables \ref{tab:c4}--\ref{tab:c5}): for these DGPs, only the first two components capture significant deviations. As a result, omnibus tests, which assign higher weights to the leading components, are more powerful than smooth tests that distribute equal weights across a larger set of components.

Second, incorporating the learning procedure generally improves power. Although the testing sample is reduced in size, selecting components based on their performance in the learning sample leads to higher power than simply using the first 5 or 10 components. This improvement is especially pronounced in DGP3-8, where the underlying alternatives are more complex and not well detected by the default leading components.

\begin{table}[!htp]
    \centering
    \caption{Estimated power for DGP6-8 at 95\%}
    \setlength{\tabcolsep}{18pt}
    \label{tab:6}
    \resizebox{\textwidth}{!}{
        \scriptsize
        \renewcommand{\arraystretch}{1.5} 
        \resizebox{0.5\textwidth}{!}{
        \begin{tabular}{@{}lllllllllllll@{}}
            \toprule
            & \multicolumn{3}{c}{$n = 300$}  \\
            \cmidrule(lr){2-4} 
            & DGP6 & DGP7 & DGP8   \\
            \midrule
            $CvM_{ch}$         & 0.948 & 0.920 & 0.932  \\
            \addlinespace
            $CvM_{es}$ & 0.656 & 0.688 & 0.644  \\
            \addlinespace
            Smooth\_5           & 0.536 & 0.488 & 0.494\\
            \addlinespace
            Smooth\_10          & 0.518 & 0.466 &0.488\\
            \addlinespace
            Smooth\_{Learn}\_5  & 0.630 & 0.620 & 0.598\\
            \addlinespace
            Smooth\_{Learn}\_2  & 0.714 & 0.726 & 0.738\\
            \addlinespace
            $CvM_{n,1}$  & 0.920 & 0.910 & 0.912\\
            \addlinespace
            $CvM_{n,2}$  & 0.924 & 0.928 & 0.938\\
            \addlinespace
            \bottomrule
        \end{tabular}
    }
    }
    \begin{tablenotes}
	$CvM_{n,j}$, $j=1,2$ are the $ j $-th component tests.  
\end{tablenotes}
\end{table}

Finally, the selection and weighting of components play a pivotal role, as illustrated by the results for DGP6-8 in Table \ref{tab:6}. In these scenarios, the first two components exhibit markedly lower p-values than the others, indicating that they capture the primary deviations from the null hypothesis. This observation motivates the construction of the $Smooth\_Learn\_2$ statistic, which aggregates only these two components. The simulation results confirm that focusing on the most informative components can substantially enhance power.

Moreover, the choice of weights assigned to each component is equally important. A comparison between the power of $Smooth\_Learn\_2$ and the individual two-component tests reveals that the weighting scheme can significantly affect test performance. While equal weighting is a practical default, identifying optimal weights remains a challenging and open problem for future research.

\subsection{An Empirical Application}
We apply the proposed component-based testing strategies to a real-world dataset from the UCI Machine Learning Repository \citep{lichman2017uci}. The ``Student Performance'' dataset contains information on student dropout, including 36 covariates for 4,424 observations. The response variable is binary, indicating dropout status. We utilize a probit model to estimate the dropout probability conditional on the covariates.

Due to the large sample size, we randomly partition the dataset into a learning sample (40\% of the data) and a testing sample (60\% of the data). We consider the first 10 components. Using the learning sample, we identify the 5 components exhibiting the smallest p-values. These selected components are then used to construct the smooth test statistic within the testing sample. For comparison, we also report results for the smooth test constructed without the learning procedure, using the first 5 and the first 10 components, respectively. Additionally, we apply the CK test and the CvM test that utilizes the characteristic function as the weighting function. The CvM test with Escanciano's projection weighting is not included, as it is rather time consuming for this dataset.

The empirical results are summarized in Table \ref{tab:7}. The component-based test statistics consistently reject the null hypothesis, exhibiting p-values effectively equal to zero. In contrast, the non-component-based tests fail to reject the null, with p-values close to unity. This outcome highlights the superior power performance of the proposed component-based methods relative to the non-component alternatives for this dataset. Interestingly, the learning procedure offered no significant improvement to the smooth test's power. As indicated by the component analysis in Table \ref{tab:8}, the primary deviation signals appear concentrated within the first 10 components, thus limiting the potential benefit of the component selection step.

\begin{table}[!htp]
    \centering
    \caption{Empirical Study Results}
    \label{tab:7}
    \scriptsize 
    \renewcommand{\arraystretch}{1.5} 
    \setlength{\tabcolsep}{6pt} 
    \begin{tabular*}{\textwidth}{@{\extracolsep{\fill}}lc@{}} 
        \toprule
        Test Statistic & p-value \\
        \midrule
        CK         & 0.972 \\
        \addlinespace
        $CvM_{ch}$         & 1.000 \\
        \addlinespace
        Smooth\_5           & 0.000 \\
        \addlinespace
        Smooth\_10          & 0.000 \\
        \addlinespace
        Smooth\_{Learn}\_5  & 0.000 \\
        \addlinespace
        \bottomrule
    \end{tabular*}
    \begin{tablenotes}
       \scriptsize 
        The Smooth\_k measure is constructed by considering the first k components with equal weights. For learning, 40\% of the data is used. The Smooth\_Learn\_k measure uses the largest k components identified in the learning phase on the remaining 60\% of the data.
    \end{tablenotes}
\end{table}

\begin{table}[!htp]
    \centering
    \caption{Component Test p-values (Learning Sample)}
    \label{tab:8}
    \scriptsize 
    \renewcommand{\arraystretch}{1.5} 
    \setlength{\tabcolsep}{6pt} 
    \begin{tabular*}{\textwidth}{@{\extracolsep{\fill}}lc@{}} 
        \toprule
        Component Test Statistic ($CvM_{n,k}$) & p-value  \\
        \midrule
        $CvM_{n,1}$        & 0.020 \\
        \addlinespace
        $CvM_{n,2}$         & 0.000 \\
        \addlinespace
        $CvM_{n,3}$         & 0.000 \\
        \addlinespace
        $CvM_{n,4}$         & 0.000 \\
        \addlinespace
        $CvM_{n,5}$         & 0.000 \\
        \addlinespace
        $CvM_{n,6}$         & 0.000 \\
        \addlinespace
        $CvM_{n,7}$         & 0.000 \\
        \addlinespace
        $CvM_{n,8}$         & 0.000 \\
        \addlinespace
        $CvM_{n,9}$         & 0.000 \\
        \addlinespace
        $CvM_{n,10}$        & 0.000 \\
        \addlinespace
        \bottomrule
    \end{tabular*}
    \begin{tablenotes}
       \scriptsize 
        P-values for the first 10 component tests.
    \end{tablenotes}
\end{table}

\section {Conclusion}

This paper focuses on two central questions in the context of testing conditional distributions: (i) whether omnibus statistics can be decomposed into a set of individual components, each capturing a distinct aspect of model fit; and (ii) how to identify which components are most informative about departures from the null hypothesis. A third, related question—how to optimally combine these components into a new, more powerful test statistic—is left for future research.

We address the first two questions by developing a conditional principal component analysis (PCA) of the omnibus test statistic. The resulting components provide a principled basis for constructing more informative specification tests. To determine the relative importance of each component, we propose a simple, data-driven learning procedure: we compute directional test statistics for each component and use bootstrap p-values to assess their significance. Components with small p-values are interpreted as carrying strong signals of model misspecification, while those with large p-values are likely to reflect noise.

Ideally, one would assign weights to each component proportional to the strength of its signal. However, determining optimal weights is a challenging problem that lies beyond the scope of this paper. As a practical solution, we adopt equal weighting of the selected components, which, despite its simplicity, yields substantial improvements in testing power in our empirical studies.

\newpage
\vskip 0.2in
\bibliography{reference}

\begin{thebibliography}{51}
\newcommand{\enquote}[1]{``#1''}
\providecommand{\natexlab}[1]{#1}
\providecommand{\url}[1]{\texttt{#1}}
\providecommand{\urlprefix}{URL }
\expandafter\ifx\csname urlstyle\endcsname\relax
  \providecommand{\doi}[1]{doi:\discretionary{}{}{}#1}\else
  \providecommand{\doi}{doi:\discretionary{}{}{}\begingroup \urlstyle{rm}\Url}\fi
\providecommand{\selectlanguage}[1]{\relax}

\bibitem[{Amengual et~al.(2020)Amengual, Carrasco, and Sentana}]{amengual2020testing}
Amengual, Dante, Marine Carrasco, and Enrique Sentana (2020). \enquote{Testing distributional assumptions using a continuum of moments.} \emph{Journal of Econometrics}, 218(2), 655--689.

\bibitem[{Andrews(1997)}]{andrews1997conditional}
Andrews, Donald~WK (1997). \enquote{A conditional Kolmogorov test.} \emph{Econometrica: Journal of the Econometric Society}, pp. 1097--1128.

\bibitem[{Bai(2003)}]{bai2003testing}
Bai, Jushan (2003). \enquote{Testing parametric conditional distributions of dynamic models.} \emph{Review of Economics and Statistics}, 85(3), 531--549.

\bibitem[{Bickel et~al.(1993)Bickel, Klaassen, Bickel, Ritov, Klaassen, Wellner, and Ritov}]{bickel1993efficient}
Bickel, Peter~J, Chris~AJ Klaassen, Peter~J Bickel, Ya’acov Ritov, J~Klaassen, Jon~A Wellner, and YA'Acov Ritov (1993). \emph{Efficient and adaptive estimation for semiparametric models}, vol.~4. Springer.

\bibitem[{Bickel and Wichura(1971)}]{bickel1971convergence}
Bickel, Peter~J and Michael~J Wichura (1971). \enquote{Convergence criteria for multiparameter stochastic processes and some applications.} \emph{The Annals of Mathematical Statistics}, 42(5), 1656--1670.

\bibitem[{Bierens(1982)}]{bierens1982consistent}
Bierens, Herman~J (1982). \enquote{Consistent model specification tests.} \emph{Journal of Econometrics}, 20(1), 105--134.

\bibitem[{Bierens(1990)}]{bierens1990consistent}
Bierens, Herman~J (1990). \enquote{A consistent conditional moment test of functional form.} \emph{Econometrica: Journal of the Econometric Society}, pp. 1443--1458.

\bibitem[{Bierens and Ploberger(1997)}]{bierens1997asymptotic}
Bierens, Herman~J and Werner Ploberger (1997). \enquote{Asymptotic theory of integrated conditional moment tests.} \emph{Econometrica: Journal of the Econometric Society}, pp. 1129--1151.

\bibitem[{Bierens and Wang(2012)}]{bierens2012integrated}
Bierens, Herman~J and Li~Wang (2012). \enquote{Integrated conditional moment tests for parametric conditional distributions.} \emph{Econometric Theory}, 28(2), 328--362.

\bibitem[{Billingsley(2013)}]{billingsley2013convergence}
Billingsley, Patrick (2013). \emph{Convergence of probability measures}. John Wiley \& Sons.

\bibitem[{Dai et~al.(2022)Dai, Song, and Xiao}]{song}
Dai, Shengtao, Xiaojun Song, and Zhijie Xiao (2022). \enquote{Specification Analysis of Conditional Distribution Functions.} \emph{Working Paper}.

\bibitem[{Delgado and Manteiga(2001)}]{delgado2001significance}
Delgado, Miguel~A and Wenceslao~Gonz{\'a}lez Manteiga (2001). \enquote{Significance testing in nonparametric regression based on the bootstrap.} \emph{The Annals of Statistics}, 29(5), 1469--1507.

\bibitem[{Delgado and Stute(2008)}]{delgado2008distribution}
Delgado, Miguel~A and Winfried Stute (2008). \enquote{Distribution-free specification tests of conditional models.} \emph{Journal of Econometrics}, 143(1), 37--55.

\bibitem[{Durbin(1973)}]{durbin1973weak}
Durbin, James (1973). \enquote{Weak convergence of the sample distribution function when parameters are estimated.} \emph{The Annals of Statistics}, pp. 279--290.

\bibitem[{Durbin and Knott(1972)}]{durbin1972components}
Durbin, James and Martin Knott (1972). \enquote{Components of Cram{\'e}r--von Mises statistics. I.} \emph{Journal of the Royal Statistical Society: Series B (Methodological)}, 34(2), 290--307.

\bibitem[{Durbin et~al.(1975)Durbin, Knott, and Taylor}]{durbin1975components}
Durbin, James, Martin Knott, and CC~Taylor (1975). \enquote{Components of Cram{\'e}r-Von Mises Statistics. Ii.} \emph{Journal of the Royal Statistical Society: Series B (Methodological)}, 37(2), 216--237.

\bibitem[{Escanciano(2006)}]{escanciano2006consistent}
Escanciano, J~Carlos (2006). \enquote{A consistent diagnostic test for regression models using projections.} \emph{Econometric Theory}, 22(6), 1030--1051.

\bibitem[{Escanciano(2009)}]{escanciano2009lack}
Escanciano, J~Carlos (2009). \enquote{On the lack of power of omnibus specification tests.} \emph{Econometric Theory}, 25(1), 162--194.

\bibitem[{Eubank and Hart(1992)}]{eubank1992testing}
Eubank, Randall~L and Jeffrey~D Hart (1992). \enquote{Testing goodness-of-fit in regression via order selection criteria.} \emph{The annals of Statistics}, pp. 1412--1425.

\bibitem[{Eubank and LaRiccia(1992)}]{eubank1992asymptotic}
Eubank, RL and VN~LaRiccia (1992). \enquote{Asymptotic comparison of Cramer-von Mises and nonparametric function estimation techniques for testing goodness-of-fit.} \emph{The Annals of Statistics}, 20(4), 2071--2086.

\bibitem[{Fan and Li(2000)}]{fan2000consistent}
Fan, Yanqin and Qi~Li (2000). \enquote{Consistent model specification tests: Kernel-based tests versus Bierens' ICM tests.} \emph{Econometric Theory}, 16(6), 1016--1041.

\bibitem[{Grenander(1950)}]{grenander1950stochastic}
Grenander, Ulf (1950). \enquote{Stochastic processes and statistical inference.} \emph{Arkiv f{\"o}r matematik}, 1(3), 195--277.

\bibitem[{Hardle and Mammen(1993)}]{hardle1993comparing}
Hardle, Wolfgang and Enno Mammen (1993). \enquote{Comparing nonparametric versus parametric regression fits.} \emph{The Annals of Statistics}, pp. 1926--1947.

\bibitem[{Hong and White(1995)}]{hong1995consistent}
Hong, Yongmiao and Halbert White (1995). \enquote{Consistent specification testing via nonparametric series regression.} \emph{Econometrica: Journal of the Econometric Society}, pp. 1133--1159.

\bibitem[{Horowitz and H{\"a}rdle(1994)}]{horowitz1994testing}
Horowitz, Joel~L and Wolfgang H{\"a}rdle (1994). \enquote{Testing a parametric model against a semiparametric alternative.} \emph{Econometric theory}, 10(5), 821--848.

\bibitem[{Huber(1985)}]{huber1985projection}
Huber, Peter~J (1985). \enquote{Projection pursuit.} \emph{The annals of Statistics}, pp. 435--475.

\bibitem[{Inglot and Ledwina(1996)}]{inglot1996asymptotic}
Inglot, Tadeusz and Teresa Ledwina (1996). \enquote{Asymptotic optimality of data-driven Neyman's tests for uniformity.} \emph{The Annals of Statistics}, 24(5), 1982--2019.

\bibitem[{Janssen(2000)}]{janssen2000global}
Janssen, Arnold (2000). \enquote{Global power functions of goodness of fit tests.} \emph{The Annals of Statistics}, 28(1), 239--253.

\bibitem[{Jennrich(1969)}]{jennrich1969asymptotic}
Jennrich, Robert~I (1969). \enquote{Asymptotic properties of non-linear least squares estimators.} \emph{The Annals of Mathematical Statistics}, 40(2), 633--643.

\bibitem[{Kac and Siegert(1947)}]{kac1947explicit}
Kac, Marc and AJF Siegert (1947). \enquote{An explicit representation of a stationary Gaussian process.} \emph{The Annals of Mathematical Statistics}, 18(3), 438--442.

\bibitem[{Kallenberg and Ledwina(1997)}]{kallenberg1997data}
Kallenberg, Wilbert~CM and Teresa Ledwina (1997). \enquote{Data-driven smooth tests when the hypothesis is composite.} \emph{Journal of the American Statistical Association}, 92(439), 1094--1104.

\bibitem[{Khmaladze(1982)}]{khmaladze1982martingale}
Khmaladze, Estate~V (1982). \enquote{Martingale approach in the theory of goodness-of-fit tests.} \emph{Theory of Probability \& Its Applications}, 26(2), 240--257.

\bibitem[{Khmaladze(1993)}]{khmaladze1993goodness}
Khmaladze, Estate~V (1993). \enquote{Goodness of fit problem and scanning innovation martingales.} \emph{The Annals of Statistics}, pp. 798--829.

\bibitem[{Lavergne and Vuong(2000)}]{lavergne2000nonparametric}
Lavergne, Pascal and Quang Vuong (2000). \enquote{Nonparametric significance testing.} \emph{Econometric Theory}, 16(4), 576--601.

\bibitem[{Ledwina(1994)}]{ledwina1994data}
Ledwina, Teresa (1994). \enquote{Data-driven version of Neyman's smooth test of fit.} \emph{Journal of the American Statistical Association}, 89(427), 1000--1005.

\bibitem[{Lichman(2017)}]{lichman2017uci}
Lichman, M (2017). \enquote{UCI Machine Learning Repository. School of Information and Computer Science, University of California, Irvine, CA (2013).} \emph{URL http://archive. ics. uci. edu/ml}.

\bibitem[{Neyman(1937)}]{neyman1937smooth}
Neyman, Jerzy (1937). \enquote{Smooth test for goodness of fit.} \emph{Scandinavian Actuarial Journal}, 1937(3-4), 149--199.

\bibitem[{Pollard(2012)}]{pollard2012convergence}
Pollard, David (2012). \emph{Convergence of stochastic processes}. Springer Science \& Business Media.

\bibitem[{Rosenblatt(1952)}]{rosenblatt1952remarks}
Rosenblatt, Murray (1952). \enquote{Remarks on a multivariate transformation.} \emph{The annals of mathematical statistics}, 23(3), 470--472.

\bibitem[{Schoenfeld(1980)}]{schoenfeld1980tests}
Schoenfeld, David (1980). \enquote{Tests based on linear combinations of the orthogonal components of the Cram{\'e}r-von Mises statistic when parameters are estimated.} \emph{The Annals of Statistics}, pp. 1017--1022.

\bibitem[{Schoenfeld(1977)}]{schoenfeld1977asymptotic}
Schoenfeld, David~A (1977). \enquote{Asymptotic properties of tests based on linear combinations of the orthogonal components of the Cramer-von Mises statistic.} \emph{The Annals of Statistics}, pp. 1017--1026.

\bibitem[{Stinchcombe and White(1998)}]{stinchcombe1998consistent}
Stinchcombe, Maxwell~B and Halbert White (1998). \enquote{Consistent specification testing with nuisance parameters present only under the alternative.} \emph{Econometric theory}, 14(3), 295--325.

\bibitem[{Stute et~al.(2008)Stute, Xu, and Zhu}]{stute2008model}
Stute, W, WL~Xu, and LX~Zhu (2008). \enquote{Model diagnosis for parametric regression in high-dimensional spaces.} \emph{Biometrika}, 95(2), 451--467.

\bibitem[{Stute(1993)}]{stute1993consistent}
Stute, Winfried (1993). \enquote{Consistent estimation under random censorship when covariables are present.} \emph{Journal of Multivariate Analysis}, 45(1), 89--103.

\bibitem[{Stute(1997)}]{stute1997nonparametric}
Stute, Winfried (1997). \enquote{Nonparametric model checks for regression.} \emph{The Annals of Statistics}, pp. 613--641.

\bibitem[{Stute and Anh(2012)}]{stute2012principal}
Stute, Winfried and TL~Anh (2012). \enquote{Principal component analysis of martingale residuals.} \emph{J. Indian Statist. Assoc}, 50(1-2), 263--276.

\bibitem[{Stute et~al.(1998)Stute, Thies, and Zhu}]{stute1998model}
Stute, Winfried, Silke Thies, and Li-Xing Zhu (1998). \enquote{Model checks for regression: an innovation process approach.} \emph{The Annals of Statistics}, 26(5), 1916--1934.

\bibitem[{Stute and Zhu(2002)}]{stute2002model}
Stute, Winfried and Li-Xing Zhu (2002). \enquote{Model checks for generalized linear models.} \emph{Scandinavian Journal of Statistics}, 29(3), 535--545.

\bibitem[{Zheng(1996)}]{zheng1996consistent}
Zheng, John~Xu (1996). \enquote{A consistent test of functional form via nonparametric estimation techniques.} \emph{Journal of Econometrics}, 75(2), 263--289.

\bibitem[{Zheng(2000)}]{zheng2000consistent}
Zheng, John~Xu (2000). \enquote{A consistent test of conditional parametric distributions.} \emph{Econometric Theory}, 16(5), 667--691.

\bibitem[{Zheng(2012)}]{zheng2012testing}
Zheng, Xu (2012). \enquote{Testing parametric conditional distributions using the nonparametric smoothing method.} \emph{Metrika}, 75, 455--469.

\end{thebibliography}

\newpage

\appendix

\section{Proofs}
\noindent\textbf{Proof of Proposition 1:}

Since $\{f_j(\cdot,x)\}_{j=1}^{\infty}$ are the eigenfunctions, the Karhunen-Lo\`eve representation of $M_i(y)$ process is thereby
$${M_i(y)}=\sum_{j=1}^{\infty}{\mu_j}^{1/2}{z}_{ij}f_j(y,X_i)~~a.s.,~ i=1,\cdots,n$$
where
\begin{align*}
z_{ij}  := & {\mu_j}^{-1/2}\langle M_i,f_j(\cdot , X_i)\rangle_{X_i}\\
= & {\mu_j}^{-1/2}\int_{\mathbb{R}}M_i(y)f_j(y,X_i)T(dy,X_i). 
\end{align*} 
The $z_{ij}$ is the $j$'th normalized PC of ${M_i(y)}$ conditional on $X_i$. 

In fact, in this distribution scenario, each PC $z_{ij}$ takes a more concise and explicit form. By applying integration by parts to the above integral, we have
\begin{align*}
z_{ij} & =\int_{\mathbb{R}}{g}_j(s,X_i)d{M}_i(s) = g_j(Y_i,X_i).	
\end{align*}
From the orthogonal properties of PCs, they have a conditional zero mean and unit variance, and are uncorrelated with each other given $X$. 

After obtaining the individual PCs, the next step involves summing them up in the same way that $R_n(y,v)$ sums up $M_i(y)$. By substituting the conditional PCD into equation (2.1), we have the decomposition of $R_n$ as
\begin{eqnarray*}
	R_n(y,v)  & = & n^{-1/2}\sum_{i=1}^{n} h(v,X_i)\left( \sum_{j=1}^{\infty}{\mu_j}^{1/2}f_j(y,X_i)g_j(Y_i,X_i)\right)\\
	& = & \sum_{j=1}^{\infty}{\mu_j}^{1/2}\left[n^{-1/2}\sum_{i=1}^{n} h(v,X_i)f_j(y,X_i)g_j(Y_i,X_i)\right],
\end{eqnarray*}
which is precisely what Proportion 1 shows.\\

\noindent\textbf{Proof of Theorem 1:} 

\noindent Note that each $f_j$ and $g_j$ are bounded and differentiable. The weak convergence of $c_{n,j}$ follows from functional CLT. The independence between $c_{\infty,j}$ and $c_{\infty,h}$ comes from the Gaussian property and conditional uncorrelation between $z_{ij}$ and $z_{ih}$. \\

\noindent\textbf{Proof of Theorem 2 and 3:} 

We first show that $\hat{c}_{n,j}(y,v)$ is asymptotically equivalent to the process $\tilde{c}_{n,j}(y,v)$, which is defined as
$$ 	
\tilde{c}_{n,j}(y,v)
:=  c_{n,j}(y,v) -  A_j(y,v) n^{-1/2}\sum_{i=1}^{n} l(X_i,Y_i,\theta_0) , 
$$
where
$$A_j(y,v)={\mu_j}^{-1/2}\mathbb{E}\left[h(v,X){f}_j(y,X)\int_{\mathbb{R}}F_{\theta}(y|
X, \theta_0)f_j(y,X)T(dy,X)  \right]$$
with $F_{\theta}$ denoting the partial derivative w.r.t. $\theta$. 

First note that the sine and cosin functions are bounded and differentiable. Since $\hat{\theta}$ is root n-consistent, $\hat{f}_j$ and $\hat{g}_j$ are uniformly consistent. 

We can write
\begin{align*}   	
	& \hat{c}_{n,j}(y,v)- n^{-1/2}\sum_{i=1}^{n} h(v,X_i){f}_j(t,X_i){g}_j(Y_i,X_i)\\
	=& \hat{c}_{n,j}(y,v)-
	n^{-1/2}\sum_{i=1}^{n} \int_{0}^{\infty}h(v,X_i)\hat{f}_j(t,X_i)\hat{g}_j(s,X_i)dM_i(s)  \\
	+& n^{-1/2}\sum_{i=1}^{n} \int_{0}^{\infty}h(v,X_i)\hat{f}_j(t,X_i)\hat{g}_j(s,X_i)dM_i(s)\\
    - &n^{-1/2}\sum_{i=1}^{n} \int_{0}^{\infty}h(v,X_i){f}_j(t,X_i){g}_j(s,X_i)dM_i(s) .
\end{align*}
For the first differnce, take the Taylor's expansion of $\hat{c}_{n,j}(y,v)$ and then we have
\begin{eqnarray*}   	
	\hat{c}_{n,j}(y,v)
	&=&  n^{-1/2}\sum_{i=1}^{n} \int_{-\infty}^{\infty}h(v,X_i)\hat{f}_j(t,X_i)\hat{g}_j(s,X_i)dM_i(s)\\
	&& -{\mu_j}^{-1/2}n^{-1}\sum_{i=1}^{n}h(v,X)\hat{f}_j(y,X)\int_{-\infty}^{\infty}F_{\theta}(y\mid X, \theta_0)\hat{f}_j(y,X)\hat{T}(dy,X) \\
	&&~ \times n^{-1/2}\sum_{i=1}^{n} l(X_i,Y_i,\theta_0)+ o_p(1)\\
	&=&  n^{-1/2}\sum_{i=1}^{n} \int_{-\infty}^{\infty}h(v,X_i)\hat{f}_j(t,X_i)\hat{g}_j(s,X_i)dM_i(s)\\
	&& -A_j(y,v)  n^{-1/2}\sum_{i=1}^{n} l(X_i,Y_i,\theta_0)+ o_p(1)
\end{eqnarray*}

To analyze the second difference, let us define
$$K(s) = s - s^2.$$
We consider the mean squared expectation:
\begin{align*}
    & \mathbb{E} \left[ n^{-1/2} \sum_{i=1}^{n} \int_{-\infty}^{\infty} \left( h(v, X_i) \hat{f}_j(y, X_i) \hat{g}_j(s, X_i) - h(v, X_i) f_j(y, X_i) g_j(s, X_i) \right) dM_i(s) \right]^2 \\
    &= \mathbb{E} \left[ n^{-1/2} \sum_{i=1}^{n} \int_{-\infty}^{\infty} \left( h(v, X_i) \hat{f}_j(y, X_i) \hat{g}_j(s, X_i) - h(v, X_i) f_j(y, X_i) g_j(s, X_i) \right)^2 K(T(ds, X_i)) \right] \\
    &= \int_{-\infty}^{\infty} \mathbb{E} \left[ \left( h(v, X_i) \hat{f}_j(y, X_i) \hat{g}_j(s, X_i) - h(v, X_i) f_j(y, X_i) g_j(s, X_i) \right)^2 K(T(ds, X_i)) \right] \\
    &\to 0.
\end{align*}

Hence the second difference is negligible, i.e.,
$$n^{-1/2}\sum_{i=1}^{n} \int_{-\infty}^{\infty}\Big(h(v,X_i)\hat{f}_j(y,X_i)\hat{g}_j(s,X_i)-h(v,X_i){f}_j(y,X_i){g}_j(s,X_i)\Big)dM_i(s)=o_p(1).$$
Thus,
\begin{eqnarray*}   	
	\hat{c}_{n,j}(y,v)
	=  n^{-1/2}\sum_{i=1}^{n} h(v,X_i){f}_j(t,X_i){g}_j(Y_i,X_i) -A_j(y,v)  n^{-1/2}\sum_{i=1}^{n} l(X_i,Y_i,\theta_0)+ o_p(1).
\end{eqnarray*} 
The weak convergence of $	{c}_{n,j}(y,v)$ and $	\hat{c}_{n,j}(y,v)$ is then easy to obtain.

\bigskip
\nocite{*}

\section{Component Test Results}
This appendix documents size and power results for different components. These results indicate features of the underlying testing problems. 

\begin{table}[!htp]
    \centering
    \caption{Component Test Results }
    \setlength{\tabcolsep}{18pt}
    \label{tab:c0}
    \resizebox{\textwidth}{!}{
        \scriptsize
        \renewcommand{\arraystretch}{1.5} 
        
        \begin{tabular}{@{}lllllllllllll@{}}
            \toprule
            & \multicolumn{3}{c}{DGP0} && \multicolumn{3}{c}{DGP1} \\
            \cmidrule(lr){2-4} \cmidrule(lr){6-8}
            & $n=100$ & $200$ & $300$  && $n=100$ & $200$ & $300$ \\
            \midrule
            $CvM_{n,1}$         & 0.098 & 0.034 & 0.042 && 0.972 & 0.982 & 0.986 \\
            \addlinespace 
            $CvM_{n,2}$         & 0.082 & 0.054 & 0.054 && 0.998 & 1.000 & 1.000 \\
            \addlinespace
            $CvM_{n,3}$         & 0.064 & 0.076 & 0.056 && 0.304 & 0.296 & 0.536 \\
            \addlinespace
            $CvM_{n,4}$         & 0.082 & 0.060 & 0.066 && 0.946 & 1.000 & 1.000 \\
            \addlinespace
            $CvM_{n,5}$         & 0.052 & 0.046 & 0.056 && 0.404 & 0.626 & 0.822 \\
            \addlinespace
            $CvM_{n,6}$         & 0.076 & 0.052 & 0.060 && 0.882 & 0.994 & 1.000 \\
            \addlinespace
            $CvM_{n,7}$         & 0.056 & 0.072 & 0.054 && 0.472 & 0.738 & 0.864 \\
            \addlinespace
            $CvM_{n,8}$         & 0.084 & 0.056 & 0.076 && 0.828 & 0.976 & 1.000 \\
            \addlinespace
            $CvM_{n,9}$         & 0.060 & 0.062 & 0.036 && 0.414 & 0.736 & 0.904\\
            \addlinespace
            $CvM_{n,10}$      & 0.094 & 0.028 & 0.062 && 0.750 & 0.944 & 1.000\\
            \addlinespace
            \bottomrule
        \end{tabular}
    }
    \begin{tablenotes}
        $CvM_{n,j}$, $j=1,2,\ldots,10$ are tests using the $ j $-th component.  
    \end{tablenotes}
\end{table}

\begin{table}[!htp]
    \centering
    \caption{Component Test Results}
    \setlength{\tabcolsep}{18pt}
    \label{tab:c1}
    \resizebox{\textwidth}{!}{
        \scriptsize
        \renewcommand{\arraystretch}{1.5} 
        
        \begin{tabular}{@{}lllllllllllll@{}}
            \toprule
            & \multicolumn{3}{c}{DGP2} && \multicolumn{3}{c}{DGP3} \\
            \cmidrule(lr){2-4} \cmidrule(lr){6-8}
            & $n=100$ & $200$ & $300$  && $n=100$ & $200$ & $300$ \\
            \midrule
            $CvM_{n,1}$         & 0.996 & 1.000 & 1.000 && 0.064 & 0.064 & 0.070 \\
            \addlinespace 
            $CvM_{n,2}$         & 0.890 & 0.968 & 0.998 && 0.008 & 0.014 & 0.016 \\
            \addlinespace
            $CvM_{n,3}$         & 0.676 & 0.978 & 1.000 && 0.176 & 0.306 & 0.392 \\
            \addlinespace
            $CvM_{n,4}$         & 0.632 & 0.792 & 0.924 && 0.042 & 0.058 & 0.070 \\
            \addlinespace
            $CvM_{n,5}$         & 0.228 & 0.326 & 0.324 && 0.128 & 0.200 & 0.288 \\
            \addlinespace
            $CvM_{n,6}$         & 0.470 & 0.714 & 0.894 && 0.084 & 0.172 & 0.250 \\
            \addlinespace
            $CvM_{n,7}$         & 0.102 & 0.120 & 0.062 && 0.068 & 0.100 & 0.172 \\
            \addlinespace
            $CvM_{n,8}$         & 0.310 & 0.578 & 0.828 && 0.120 & 0.224 & 0.316 \\
            \addlinespace
            $CvM_{n,9}$         & 0.082 & 0.072 & 0.082 && 0.082 & 0.082 & 0.096\\
            \addlinespace
            $CvM_{n,10}$      & 0.228 & 0.442 & 0.706 && 0.118 & 0.156 & 0.300\\
            \addlinespace
            \bottomrule
        \end{tabular}
    }
    \begin{tablenotes}
        $CvM_{n,j}$, $j=1,2,\ldots,10$ are tests using the $ j $-th component.  
    \end{tablenotes}
\end{table}

\begin{table}[!htp]
    \centering
    \caption{Component Test Results}
    \setlength{\tabcolsep}{18pt}
    \label{tab:c3}
    \resizebox{\textwidth}{!}{
        \scriptsize
        \renewcommand{\arraystretch}{1.5} 
        
        \begin{tabular}{@{}lllllllllllll@{}}
            \toprule
            & \multicolumn{3}{c}{DGP4} && \multicolumn{3}{c}{DGP5} \\
            \cmidrule(lr){2-4} \cmidrule(lr){6-8}
            & $n=100$ & $200$ & $300$  && $n=100$ & $200$ & $300$ \\
            \midrule
            $CvM_{n,1}$         & 0.064 & 0.064 & 0.070 && 0.100 & 0.030 & 0.074 \\
            \addlinespace 
            $CvM_{n,2}$         & 0.008 & 0.014 & 0.016 && 0.004 & 0.008 & 0.022 \\
            \addlinespace
            $CvM_{n,3}$         & 0.176 & 0.306 & 0.392 && 0.180 & 0.316 & 0.368 \\
            \addlinespace
            $CvM_{n,4}$         & 0.042 & 0.058 & 0.070 && 0.038 & 0.066 & 0.082 \\
            \addlinespace
            $CvM_{n,5}$         & 0.128 & 0.200 & 0.288 && 0.134 & 0.194 & 0.274\\
            \addlinespace
            $CvM_{n,6}$         & 0.084 & 0.172 & 0.250 && 0.096 & 0.138 & 0.262 \\
            \addlinespace
            $CvM_{n,7}$         & 0.068 & 0.100 & 0.172 && 0.068 & 0.120 & 0.124 \\
            \addlinespace
            $CvM_{n,8}$         & 0.120 & 0.224 & 0.316 && 0.116 & 0.202 & 0.310 \\
            \addlinespace
            $CvM_{n,9}$         & 0.082 & 0.082 & 0.096 && 0.066 & 0.062 & 0.084\\
            \addlinespace
            $CvM_{n,10}$      & 0.118 & 0.156 & 0.300 && 0.114 & 0.176 & 0.248\\
            \addlinespace
            \bottomrule
        \end{tabular}
    }
    \begin{tablenotes}
        $CvM_{n,j}$, $j=1,2,\ldots,10$ are tests using the $ j $-th component.  
    \end{tablenotes}
\end{table}

\begin{table}[!htp]
    \centering
    \caption{Component Test Results}
    \setlength{\tabcolsep}{18pt}
    \label{tab:c4}
    \resizebox{\textwidth}{!}{
        \scriptsize
        \renewcommand{\arraystretch}{1.5} 
        
        \begin{tabular}{@{}lllllllllllll@{}}
            \toprule
            & \multicolumn{3}{c}{DGP6} && \multicolumn{3}{c}{DGP7} \\
            \cmidrule(lr){2-4} \cmidrule(lr){6-8}
            & $n=100$ & $200$ & $300$  && $n=100$ & $200$ & $300$ \\
            \midrule
            $CvM_{n,1}$         & 0.870 & 0.896 & 0.920 && 0.866 & 0.908 & 0.910 \\
            \addlinespace 
            $CvM_{n,2}$         & 0.742 & 0.848 & 0.924 && 0.732 & 0.856 & 0.928 \\
            \addlinespace
            $CvM_{n,3}$         & 0.022 & 0.004 & 0.000 && 0.008 & 0.004 & 0.000 \\
            \addlinespace
            $CvM_{n,4}$         & 0.172 & 0.060 & 0.092 && 0.152 & 0.076 & 0.054 \\
            \addlinespace
            $CvM_{n,5}$         & 0.020 & 0.000 & 0.002 && 0.014 & 0.006 & 0.004\\
            \addlinespace
            $CvM_{n,6}$         & 0.118 & 0.218 & 0.440 && 0.132 & 0.184 & 0.368 \\
            \addlinespace
            $CvM_{n,7}$         & 0.010 & 0.016 & 0.044 && 0.018 & 0.004 & 0.020 \\
            \addlinespace
            $CvM_{n,8}$         & 0.060 & 0.100 & 0.128 && 0.064 & 0.076 & 0.114 \\
            \addlinespace
            $CvM_{n,9}$         & 0.006 & 0.010 & 0.014 && 0.008 & 0.018 & 0.010\\
            \addlinespace
            $CvM_{n,10}$      & 0.050 & 0.052 & 0.118 && 0.072 & 0.074 & 0.134\\
            \addlinespace
            \bottomrule
        \end{tabular}
    }
    \begin{tablenotes}
        $CvM_{n,j}$, $j=1,2,\ldots,10$ are tests using the $ j $-th component.  
    \end{tablenotes}
\end{table}

\begin{table}[!htp]
    \centering
    \caption{Component Test Results}
    \setlength{\tabcolsep}{18pt}
    \label{tab:c5}
    \resizebox{\textwidth}{!}{
        \scriptsize
        \renewcommand{\arraystretch}{1.5} 
        
        \begin{tabular}{@{}lllllllllllll@{}}
            \toprule
            & \multicolumn{3}{c}{DGP8} \\
            \cmidrule(lr){2-4} 
            & $n=100$ & $200$ & $300$  \\
            \midrule
            $CvM_{n,1}$         & 0.884 & 0.920 & 0.912 \\
            \addlinespace 
            $CvM_{n,2}$         & 0.784 & 0.876 & 0.938 \\
            \addlinespace
            $CvM_{n,3}$         & 0.024 & 0.006 & 0.002 \\
            \addlinespace
            $CvM_{n,4}$         & 0.150 & 0.068 & 0.050 \\
            \addlinespace
            $CvM_{n,5}$         & 0.010 & 0.010 & 0.002\\
            \addlinespace
            $CvM_{n,6}$         & 0.112 & 0.240 & 0.402 \\
            \addlinespace
            $CvM_{n,7}$         & 0.010 & 0.008 & 0.024 \\
            \addlinespace
            $CvM_{n,8}$         & 0.072 & 0.062 & 0.106 \\
            \addlinespace
            $CvM_{n,9}$         & 0.010 & 0.004 & 0.014\\
            \addlinespace
            $CvM_{n,10}$      & 0.052 & 0.082 & 0.172\\
            \addlinespace
            \bottomrule
        \end{tabular}
    }
    \begin{tablenotes}
        $CvM_{n,j}$, $j=1,2,\ldots,10$ are tests using the $ j $-th component.  
    \end{tablenotes}
\end{table}


\end{document}